\documentclass[12pt,twoside]{article}
\usepackage{fleqn,espcrc1}
\usepackage{graphicx,epsf,subfigure}

\newcommand{\AmS}{{\protect\the\textfont2
  A\kern-.1667em\lower.5ex\hbox{M}\kern-.125emS}}

\title{Small clusters of fermions}

\author{J. Carbonell and R. lazauskas\address{Lab. Physique Subatomique et Cosmologie,
        53 Av. Martyrs, 38026 Grenoble, France}}

\begin{document}
\maketitle

Several theoretical studies \cite{Pieper_FB17} indicate that there is no any
reasonable chance for 3n and 4n to be bound.
GANIL experiment \cite{MMM_PRC65_02} suggesting tetraneutron has not been confirmed
but these no-result had however the merit of pushing
theoreticians at work to clarify the possible existence of small n-clusters.
If the situation seems clear for A=3 and 4, it is less
well established  for a larger number of neutrons.
The search of multi-neutron resonances raises also some interest and a recent experiment
scanning the 4n continuum in the d($^8$He,$^6$ Li)4n reaction
reports a 2-3 MeV  width structure \cite{Baumel}.

In this issue it is enlightening to make a
parallel with a similar, better-known, fermionic system: the $^3$He atomic clusters.
Since  small $^3$He droplets can exist \cite{GN_PRL84_01},
should we expect
some stability islands (see figure below) in the continuum neutron states going from N=2 to N=$\infty$.
If yes, where? If not, why?
\begin{figure}[htbp]
\begin{center}\epsfxsize=10cm\epsfysize=1.5cm{\epsfbox{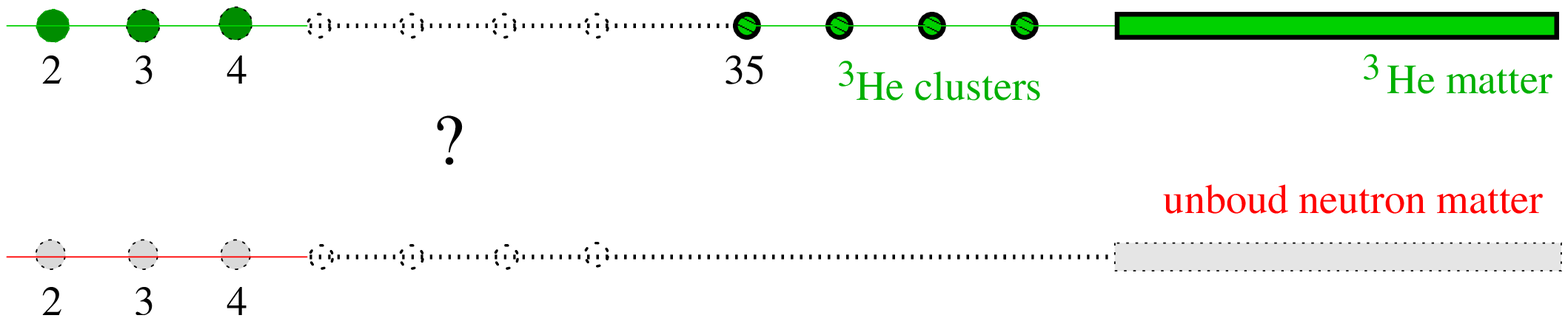}}\end{center}
\end{figure}

\vspace{-1.3cm}

\begin{figure}[h!]
\begin{minipage}[h!]{7.5cm}
\epsfxsize=7.5cm{\epsffile{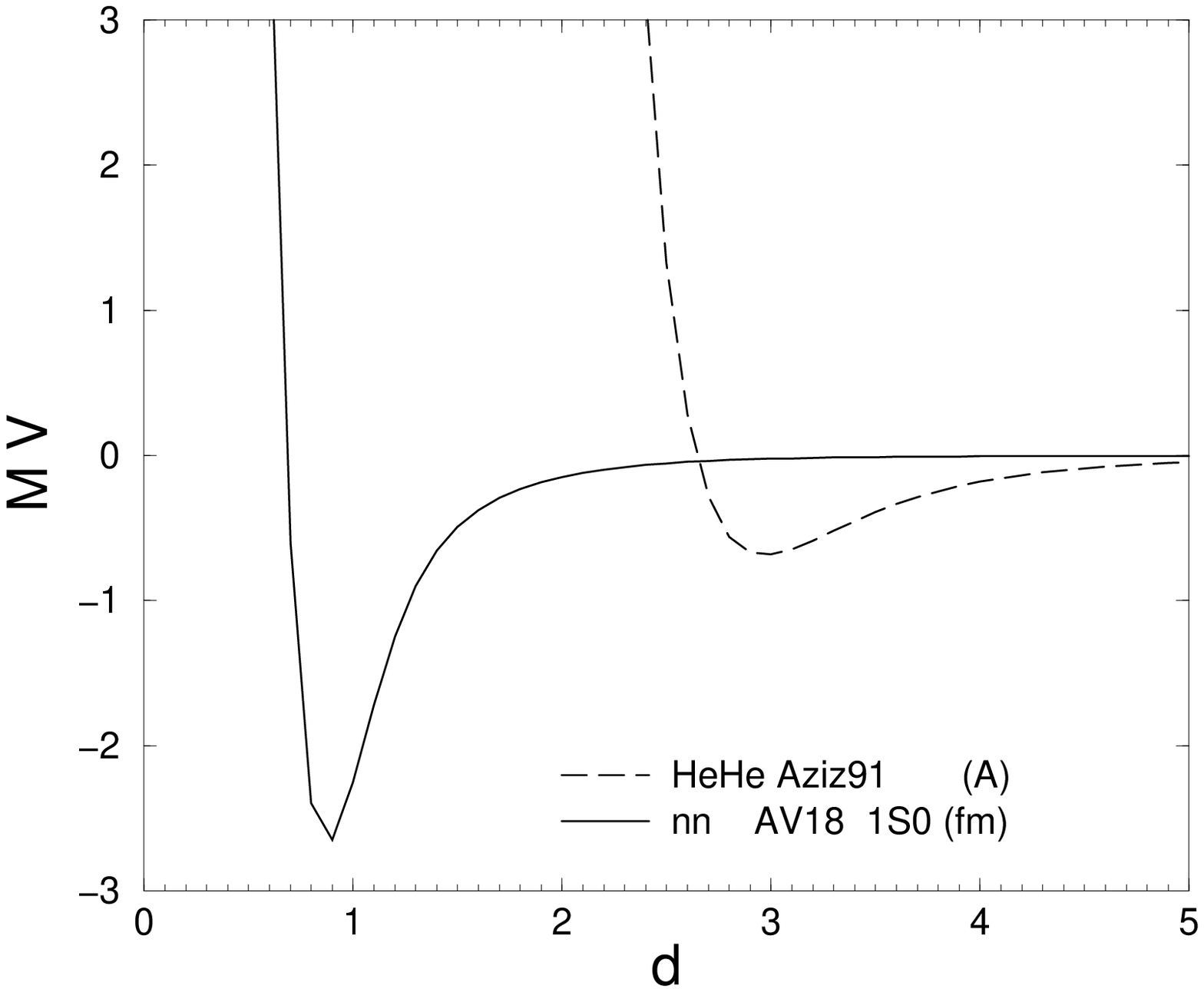}}
\vspace{-1.5cm}
\caption{Comparison between n-n  and $^3$He-$^3$He potentials.}\label{Fig_Vnn_VHeHe}
\vspace{-.5cm}
\end{minipage}
\hspace{.5cm}
\begin{minipage}[h!]{7.5cm}
\epsfxsize=7.5cm\mbox{\epsffile{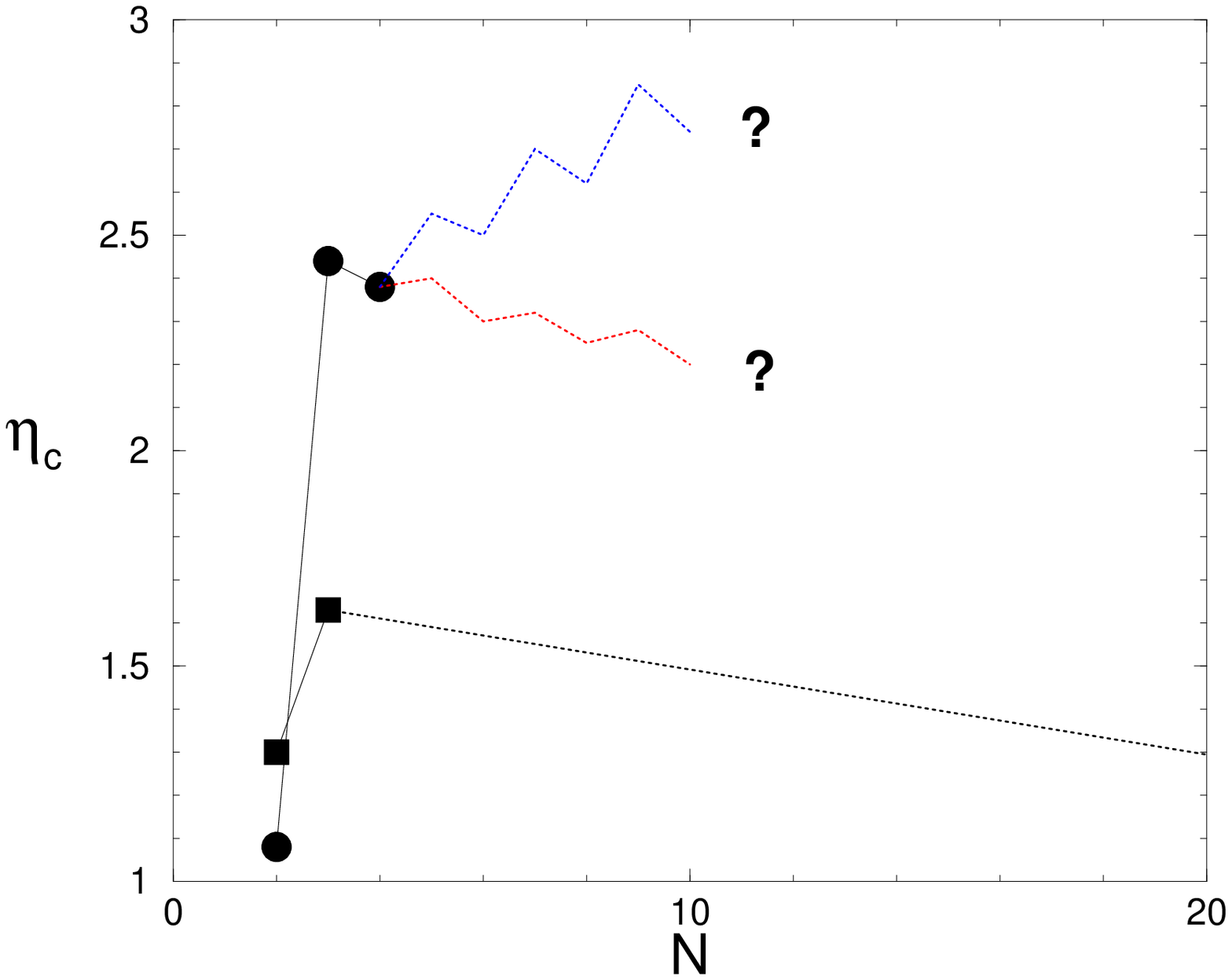}}
\vspace{-1.5cm}
\caption{Critical value of the scaling factor for $n$ (circles) and $^3$He (squares)}\label{Fig_eta_N_nn}
\vspace{-.5cm}
\end{minipage}
\end{figure}

The two-body interactions of these systems  look at first glance quite
similar.  We have compared in Fig. \ref{Fig_Vnn_VHeHe} the
$n$-$n$ AV18 \cite{AV18_PRC_95} and the $^3$He-$^3$He Aziz
\cite{Aziz_91} S-wave potentials. They have been rescaled by the
 corresponding masses ${M\over\hbar^2}$
and the resulting length units, as well as the inter-particle distance $d$ are
  $fm$ and $\AA$ respectively.
Corresponding low energy parameters are given in Table \ref{LEP}.
In the $n$-$n$ case we have also considered  Reid93
\cite{NIJ_PRC_93} and
MT13 \cite{MT_NPA_69} potentials, the latter being adjusted to reproduce the experimental
scattering length.
\begin{table}[htb]
\vspace{-.6cm}
\caption{Low energy n-n and $^3$He-$^3$He parameters.}\label{LEP}
\newcommand{\m}{\hphantom{$-$}}
\newcommand{\cc}[1]{\multicolumn{1}{c}{#1}}
\renewcommand{\tabcolsep}{0.3pc} 
\renewcommand{\arraystretch}{.9} 
\begin{tabular}{|c||cccc||c|} \cline{2-6}
 \multicolumn{1}{c||}{}   &\multicolumn{4}{|c||}{n-n(fm)}& {He-He($\AA$)} \\ \cline{2-6}
\multicolumn{1}{c||}{}     &  Av18   &   Reid & MT13   & Exp. & Aziz 91 \\\hline\hline
a     & -18.49 & -17.54 & -18.59 & $-18.59\pm0.4$& -7.24     \\
$r_0$ &        &  2.85  & 2.94   & $2.75\pm0.1$  &  13.5   \\
 $\eta_c$ &   1.08     &  1.09  & 1.10   &   &  1.30   \\\hline
\end{tabular} \hspace{0.cm}
\vspace{-.8cm}
\end{table}
None of these system supports a bound dimer ($a<0$) but
$^3$He seems less favorable to make clusters.
This can be seen by calculating
the critical values of the scaling factor $\eta_c$ introduced in the potential
$V^{(\eta)}(r)=\eta V_{nn}(r)$, which bounds a dimer.
For $n$-$n$, this value is around $\eta_c=1.08$
whereas for He-He is sensibly greater $\eta_c=1.30$ (see Table \ref{LEP}).

When examining larger systems we first consider the bosonic case.
Despite the absence of dimers, "bosonic" neutron
trimers and tetramers do exist with binding energies  $B_{n_3}\approx1$ MeV
and $B_{n_4}\approx10$ MeV.
This is however not the case for atomic $^3$He, suggesting once more
that neutron clusters
should be favored with respect to atomic $^3$He.
They all disappear when the Pauli principle is imposed,
but can reappear -- as in $^3$He -- if the number
of interacting particles is increased.
The existence of small clusters results thus from a
compromise between an attractive
pairwise interaction and the effective Pauli repulsion.

In order to study such a balance, we have investigated in several directions \cite{PhD_03}:
{\it (i)} scaling factor in $V_{nn}$, {\it (ii)} three-neutron interactions (TnI),
{\it (iii)} influence of n-n P- waves
{\it (iv)} confining the system in an harmonic oscillator (HO) trap
and {\it (iv)} "dimer"-"dimer" scattering.
It is always possible to bind 3- and 4-n states by
modifying the usual n-n and/or TnI models
but  the  violation has to be very strong, producing serious anomalies.
Thus,  in case  {\it (i)} one needs a large scaling factor $\eta_c\sim 3$.
Using ad-hoc TnI,
one gets a very compact object in which NN force becomes even repulsive.
Keeping the usual S-wave $n$-$n$ potential, the enhancement of $n$-$n$ P-waves
required to bound a 3- or 4-n system is such that they become themselves resonant!

We have confined N=2,3,4 $n$ in an HO trap
with fixed frequency $\omega$ and size parameter $b=\sqrt{\frac{\hbar}{m\omega}}$.
HO is the only external field in which "internal" and "center of
mass" energies can be properly separated.
In absence of $n$-$n$ forces the "internal" energies are known analytically
but can be obtained as well by solving the N-body "internal" problem, with
pairwise HO potential of frequency $\left({\omega\over\sqrt N}\right)$.
The effect of  $n$-$n$ interaction has been
evaluated by solving  the internal problem with
\vspace{-.2cm}
\[ V_{ij} = \frac{1}{2}\;m\;\left({\omega\over\sqrt N}\right)^2\;
r_{ij}^2 + V_{nn}(r_{ij}) \]
\vspace{-.2cm}
and  calculating the difference between the pure $HO$ energy 
and the $HO+V_{nn}$ one: $B_N=E^{(N)}_{HO}-E^{(N)}_{HO+nn}$.

\begin{table}[tbh!]
\vspace{-.6cm}
\caption{Binding energies $B_N$ of N neutrons in an HO trap with size parameter $b$.}\label{HO}
\renewcommand{\tabcolsep}{0.45pc} 
\renewcommand{\arraystretch}{.8} 
\begin{tabular}{|cc|cccc|cccc|cccc|}\cline{3-14}
\multicolumn{2}{c|}{ } &\multicolumn{4}{|c}{b=2} &\multicolumn{4}{|c}{b=3} & \multicolumn{4}{|c|}{b=4} \\ \hline
$N$ & $J^{\pi }$ & $E_{HO}^{(N)}$ & $B_{N}$ & $ \frac{B_{N}}{N}$ & $ \frac{B_{N}}{E_{HO}}$ & $E_{HO}^{(N)}$ & $B_{N}$
& $\frac{B_{N}}{N}$ & $\frac{B_{N}}{E_{HO}}$ &
$E_{HO}^{(N)}$ & $B_{N}$ & $\frac{B_{N}}{N}$ & $\frac{B_{N}}{E_{HO}}$ \\\hline\hline
2 & $0^{+}$ & 15.55 & 6.34 & 3.17 & 0.41 & 6.91 & 3.13 & 1.56 & 0.45 & 3.89 &
1.81 & 0.93 & 0.47 \\
3 & ${\frac{3}{2}}^{-}$ & 41.47 & 9.74 & 3.25 & 0.23 & 18.43 & 4.41 & 1.47 &
0.24 & 10.36 & 2.55 & 0.85 & 0.25 \\
4 & $0^{+}$ & 67.39 & 15.30 & 3.58 & 0.23 & 29.95 & 7.40 & 1.69 & 0.25 & 16.82
& 4.31 & 1.08 & 0.26 \\\hline
\end{tabular}
\vspace{-.6cm}
\end{table}

Results concerning the ground state are given in Table \ref{HO} for several values of $b$.
Some comments are in order:
{\it (i)} there is a clear indication of paring effect when going from  N=$2\to3\to4$
{\it (ii)} one has always $B_4>2B_2$, suggesting an effective attraction between dineutrons
{\it (iii)} the binding energy per particle increases when going from
N=2 to N=4 (iv) the ratio ${B_N\over E_{HO}}$ tends to a constant value
independent of $b$.
The preceding results tend to indicate that there is a  benefit when going from 2 to 4.
The main difference in respect to $^3$He has been found in the role
of P-waves. Their influence in $n$ case is attractive but very small
whereas they significantly contribute to the $^3$He binding energy ($\sim40\%$).
The reason for such a different behaviour
is the hard core radius of the corresponding
potentials, which differ by a factor 3 (see Fig. \ref{Fig_Vnn_VHeHe}). 
The centrifugal barrier is one order of magnitude smaller
in $^3$He and the effective potential is, contrary to $n$ case, still attractive
in regions where it can play a role.
This difference can be dramatic in binding larger fermion systems.

Despite the negative results quoted in \cite{Pieper_FB17},
the question of larger neutron clusters merits some attention.
At present, the strongest argument against their existence are the
mean-field results, all concluding to an unbounded infinite nuclear matter but it should be possible
to reach the same conclusion "from below", i.e. from a systematic 
study of few-neutron systems.
This can be performed by studying the N-dependence of the critical scaling factor $\eta_c$,
calculated simultaneously for $n$ and $^3$He. Our technology
allow us to reach only N=4 with the results displayed on Fig. \ref{Fig_eta_N_nn}.
The value $\eta_c^{(n)}$ makes a large jump when passing
from N=2 to N=3 but starts to decreases when going from N=3 to
N=4. Is that a pure numerical accident or, as in $^3$He, an indication 
of a descent towards the $\eta=1$ axis?.
More powerful methods could go far beyond N=4 and draw a definite conclusion
on this problem.


\begin{thebibliography}{9}
\bibitem{Pieper_FB17}  S.C. Pieper, Phys. Rev. Lett. {\bf 90} (2003) 252501; contrib. to FB17
 and refs. therein
\bibitem{MMM_PRC65_02} F.M. Marques et al, Phys. Rev. {\bf C65}, 044006 (2002).
\bibitem{Baumel}       D. Baumel, private communication
\bibitem{GN_PRL84_01}  R. Guardiola, J. Navarro, Phys. Rev. Lett. {\bf84},  1144 (2001).
\bibitem{AV18_PRC_95}  R.B. Wiringa, V.G.J. Stoks, R. Schiavilla,     Phys. Rev.  {\bf C51}  (1995) 38
\bibitem{Aziz_91}      R.A. Aziz and M.J. Slaman, J. Chem. Phys. {\bf 94} (1991) 8047
\bibitem{NIJ_PRC_93}   V.G. Stoks, R.A. Klomp, C.P. Terheggen, J.J. de Swart, 
Phys. Rev. {\bf C49} (1994)  2950
\bibitem{MT_NPA_69}    R.A. Malfliet, J. Tjon, Nucl. Phys.   {\bf A127} (1969) 161
\bibitem{PhD_03}       R. Lazauska, PhD. Universit\'e de Grenoble (2003)
\end{thebibliography}
\end{document}